\documentclass[preprint,aps,superscriptaddress,floats,showpacs]{revtex4}
\usepackage{graphicx}
\usepackage{amsmath}
\usepackage{amsfonts}
\usepackage{amssymb}
\usepackage{epsfig}
\usepackage{color}
\usepackage{bm}

\begin{document}

\title{Effective production of gammas, positrons, and photonuclear particles
from optimized electron acceleration by short laser pulses in low-density targets}

\author{M.~G.~Lobok}
\affiliation{P.~N.~Lebedev Physics Institute,
	Russian Academy of Science, Leninskii Prospect 53, Moscow 119991,
	Russia}
\affiliation{Center for Fundamental and Applied Research,
Dukhov Research Institute of Automatics (VNIIA), Moscow 127055, Russia}
\author{A.~V.~Brantov}
\affiliation{P.~N.~Lebedev Physics Institute, Russian Academy of
	Science, Leninskii Prospect 53, Moscow 119991, Russia}
\affiliation{Center for Fundamental and Applied Research,
Dukhov Research Institute of Automatics (VNIIA), Moscow 127055, Russia}
\author{V.~Yu.~Bychenkov}
\affiliation{P.~N.~Lebedev Physics Institute,
Russian Academy of Science, Leninskii Prospect 53, Moscow 119991,
Russia}
\affiliation{Center for Fundamental and Applied Research,
	Dukhov Research Institute of Automatics (VNIIA), Moscow 127055, Russia}

\begin{abstract}
Electron acceleration has been optimized based on 3D PIC simulations of a
short laser pulse interacting with low-density plasma targets to find the
pulse propagation regime that maximizes the charge of high-energy electron
bunches. This regime corresponds to laser pulse propagation in a
self-trapping mode where the diffraction divergence is balanced by the
relativistic nonlinearity such that relativistic self-focusing on the axis
does not happen and the laser beam radius stays unchanged during pulse
propagation in a plasma over many Rayleigh lengths. Such a regime occurs for
a near-critical density if the pulse length considerably exceeds both the
plasma wavelength and the pulse width. Electron acceleration occurs in a
traveling cavity filled with a high-frequency laser field and a longitudinal
electrostatic single-cycle field (``self-trapping regime''). Monte Carlo
simulations demonstrated a high electron yield that allows an efficient
production of gamma radiation, electron--positron pairs, neutrons, and even
pions from a catcher-target.
\end{abstract}


\maketitle

\section{Introduction}\label{sec1}

Electron acceleration by a laser-excited plasma wakefield with a phase
velocity close to the speed of light \cite{Tajima} is a promising source
for many applications. Excitation of a wakefield is most efficient when
the laser pulse length $L$ is of the order of the plasma wavelength,
$L=\lambda_p/2$ \cite{Esarey}. A relativistically intense laser pulse can
expel plasma electrons outward to create a traveling bare ion cavity. Such a
3D wakefield structure, which has been observed in PIC (particle-in-cell)
simulations, remains stable for laser pulses shorter than the plasma
wavelength, $L<\lambda_p$, and the laser pulse width, $L<d$ \cite{Pukhov}.
This regime, called the ``bubble'' regime is now the most used base for
electron acceleration and X-ray radiation sources.

We recently performed an optimization study to correctly define the density
and thickness of planar low-density targets maximizing the number of
high-energy electrons generated by a femtosecond laser pulse of a given
intensity. The interest in such a study is related to laser production of
electron bunches that can produce hard gamma quanta with an energy exceeding
1\,MeV suitable for radiography of thick dense samples, electron--positron
pair production, different ($\gamma$, n) reactions including neutron
generation, and even light mesons (muons and pions). A high total charge of
accelerated electron bunches is required for these applications to be of
practical significance. In this context, studies of laser interaction with
targets of a density considerably exceeding standard rarefied gas densities
(used for wakefield/bubble acceleration) \cite{Yang,Willingale,Goers} become
attractive and should be advanced still further in the direction proposed in
Ref.~\cite{Lobok}, where it was shown that ultrarelativistic laser pulses
($a_0\gg1$) of $\sim4$\,J energy propagating in a near-critical plasma
enable acceleration of a significant number of electrons (7\,nC) to an
energy $>30$\,MeV, which could interact with bremsstrahlung converter
targets and produce a desired number of gamma quanta \cite{Lobok}. The
considered acceleration regime is quite different from a standard bubble
regime with conditions opposite to Ref.~\cite{Pukhov}: $L>\lambda_p,\,d$
\cite{Lobok}. Electron acceleration occurs in a traveling nonlinear 3D
charge-separation structure as an elongated cavity (laser bullet) filled
with a laser field and not in an empty bubble \cite{Pukhov}.

Here, we advance the study of the self-trapping regime of light propagation
in a dense gas plasma to obtain the maximum yield of gammas, light
elementary particles, and photonuclear reactions. The sources of X-ray
radiation based on laser-triggered electrons have broad potential
applications including medical and biological imaging, diagnostics for
materials science, probing of dense plasmas, and security (inspection)
systems \cite{albert}. Similar sources can also be used for the strength
bench test of microchips, which is important for improving the reliability
of the components of electronics operating in space and safety systems.
Several applications of laser-based X-ray sources have so far been
demonstrated: for diagnostic radiology involving a phase-contrast imaging
technique \cite{wenz}, for inspection relying on high-contrast gamma-ray
radiography \cite{courties,brenner}, for production of medical isotopes via
photonuclear reactions in nuclear medicine \cite{luo}, for induction of
photofission \cite{reed}, and for the radiography of high-temperature
plasmas \cite{albert}. These applications require a high brightness of the
X-ray pulse generated during single laser shot. This can be achieved by
increasing the number of laser accelerated electrons, and acceleration in
the self-trapping regime could be the best choice.

We here extend our preliminary study of the self-trapping acceleration
regime \cite{Lobok}. Using 3D PIC simulations, we investigate the interplay
of the laser pulse and plasma parameters to choose the best parameter values
for maximizing the total charge of the electron bunch produced during
acceleration in the self-trapping propagation regime in a near-critical
plasma. In our numerical experiment, the generated electron bunches were
aimed at a bremsstrahlung converter target where their interaction leads to
production of gamma pulses, electron--positron pairs, neutrons from
photonuclear reaction, and more exotic particles such as light mesons,
pions, and muons. The corresponding yields were calculated using the GEANT4
code.

\section{Simulation model}

Here, we use 3D PIC simulations with the high-performance electromagnetic
code VSim (VORPAL) to study the generation of electron bunches by a
laser-generated space-charge structure from an underdense planar target
represented in the form of a plasma slab of given electron density and
thickness. Because we are interested in an electron beam suitable as a
source for deep gamma radiography and photoproduction of neutrons and
elementary particles, we choose a laser pulse that can accelerate electrons
up to a typical energy (``temperature'') exceeding 100\,MeV. This requires
laser pulses of energies considerably higher than 1\,J or sub-PW powers. For
such femtosecond pulses, we study the laser--plasma interaction with
low-density targets of different densities and thicknesses to find their
optimum values maximizing the number of accelerated electrons with energies
of practical interest.

We consider a linearly polarized laser pulse (in the $z$ direction) with the
wavelength $\lambda=2\pi c/\omega=1\,\mu$m of variable energy and a Gaussian
intensity amplitude shape in time with the FWHM duration $\tau=30$\,fs and a
Gaussian amplitude profile of the focal spot with the FWHM size $D=2R_L=
4\,\mu$m incident along the normal to the target in the $x$ direction. For
these parameters, the standard normalized laser field amplitude $a_0=eE_L/
m_e\omega c$ was varied in the range $a_0=24$ to 72, which corresponds to a
maximum laser pulse intensity of (0.8 to $7)\times10^{21}$ and a laser power
135 to 1200\,TW. The laser pulse was focused on the front side of a plasma
target consisting of electrons and heavy immovable ions. The electron
densities were in the range from a few percent of the electron critical
density ($n_c$) to a few critical densities. The target thicknesses $l$ was
varied from the pulse length $L=c\tau$ to that corresponding to almost
entire pulse depletion. The simulations were performed with a moving-window
technique with spatial grid steps $0.04\lambda\times0.1\lambda\times0.1
\lambda$ in a simulation window $X\times Y\times Z=58\lambda\times25\lambda
\times25\lambda$.
\begin{figure} [!ht]
	\centering{\includegraphics[width=16.8cm]{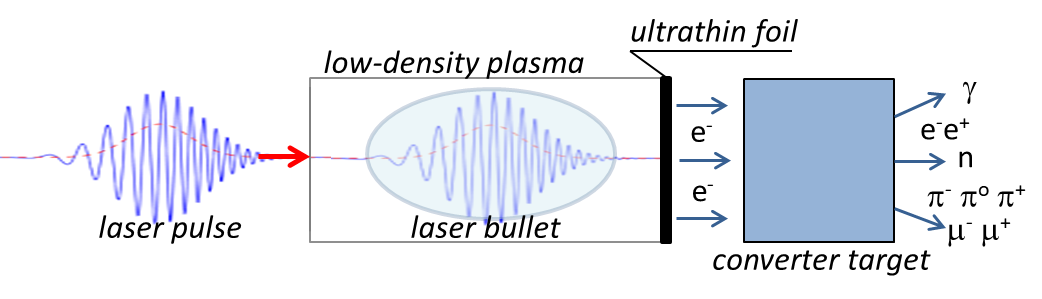}}
	\caption{Simulation laser--target layout}
	\label{fig1}
\end{figure}

In our simulations, we manipulated the parameters of a planar low-density
target to allow stable laser pulse propagation and effective generation of
electron bunches with a high average energy and high total charge. In this
study, we concentrated on studying how to accelerate as many electrons as
possible. These accelerated electrons were used as an input source for
further simulations of nuclear reactions in a converter target placed behind
the laser target. The corresponding simulations were performed with the
GEANT4 code. An ultrathin overdense plasma foil ($l=2\,\mu$m, $n_e=20n_c$)
covered the backside of the low-density target to reflect the residual part
of the laser pulse reaching the target backside; this allowed cleanly
measuring the energies of free-streaming electrons without any effect of the
transmitted laser pulse. This slab entirely isolated the laser pulse from
the high-Z converter target without affecting the accelerated electrons.
The simulation laser--target layout is shown in Fig.~\ref{fig1}.

\section{Self-trapping regime}

\begin{figure} [!ht]
	\centering{\includegraphics[width=16.4cm]{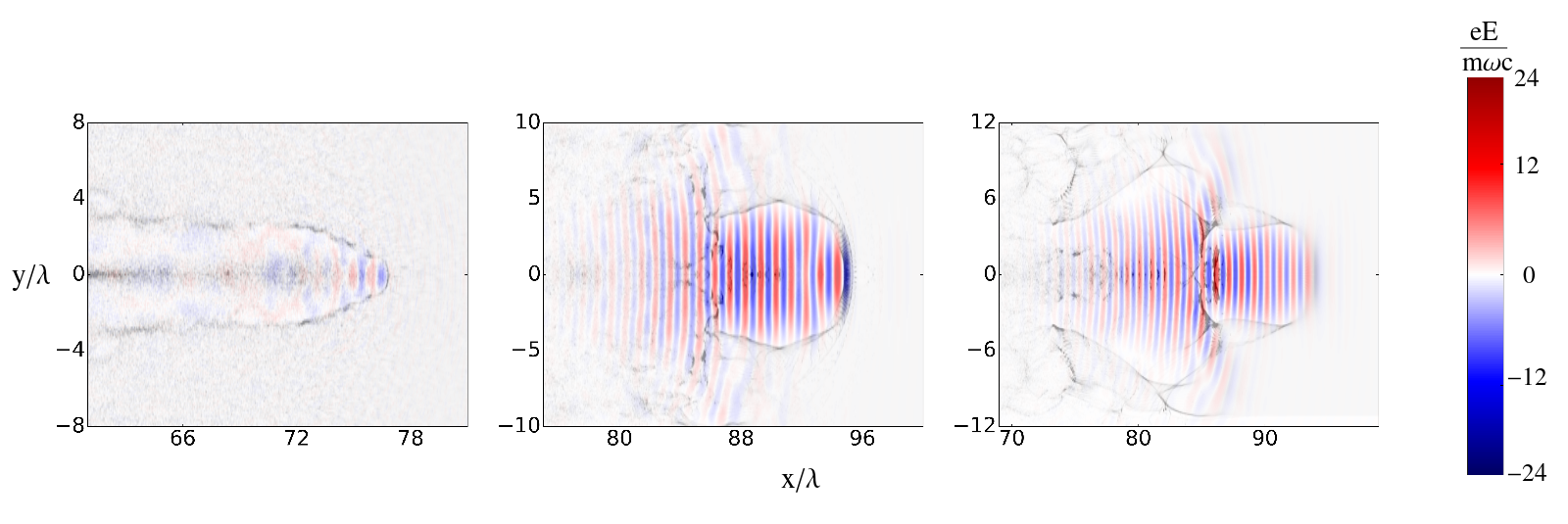}}
	\caption{Laser--plasma images for different electron densities $n_e=n_c$,
$n_e=0.1n_c$, and $n_e=0.05n_c$ (in the first, middle, and last panels) after
pulse propagation over three Rayleigh lengths within a target for
$R_L=2\lambda$ and $a_0=24$.}
	\label{fig2}
\end{figure}

Relativistic laser pulse self-focusing and self-trapping play an important
role in the stable propagation of a laser pulse over several Rayleigh
lengths needed for producing high-energy, high-current electron beams inside
the cavitated plasma cavern. The simulations performed clearly demonstrate
the importance of matching the self-consistent waveguide radius $R$ to the
electron plasma density $n_e$ for given laser power \cite{mourou,gordienko,lu}.
As a result of simulations, the proposed condition for matching $R$ to $n_e$
for indestructible pulse propagation in a relativistic plasma ($\gamma\sim
a_0$) is \cite{gordienko,lu}
\begin{equation}
R\simeq\alpha\frac{c}{\omega_p}\sqrt{a_0}=
\frac{c}{\omega}\sqrt{a_0\frac{\alpha^2n_c}{n_e}}\,,
\label{eq1}
\end{equation}
where $\omega_p$ is the electron plasma frequency ($\lambda_p=2\pi c/
\omega_p$), $n_c$ is the electron critical density for the laser frequency
$\omega$, and $\alpha$ is a numerical factor of the order of unity.
Figure~\ref{fig2} from our simulations clearly shows that only a certain
cavity radius for given plasma density is well suited for a stable 3D
soliton-like structure filled with a high-frequency laser field (see the
middle panel: $R\simeq5\lambda\simeq2(c/\omega_p)\sqrt{a_0}$, $n_e=0.1n_c$,
$a_0=24$, $R_L=2\lambda$). Here, the laser field is shown in color,
electrons are shown in gray, and the electrostatic field is shown by the
dashed curve.
We can use Eq.~(\ref{eq1}) to reformulate the matched laser cavern spot size
condition in terms of the laser power $P=E_0^2R^2c/8$ as
\begin{equation}\label{eq2}
R=\frac{c}{\omega}\sqrt{\frac{n_c}{n_e}}
\left(\frac{16\alpha^4P}{P_c}\right)^{1/6}\quad{\rm or}\quad
a_0=\left(\frac{16P}{\alpha^2P_c}\right)^{1/3},
\end{equation}
where $P_c=2(m_ec^3/r_e)(\omega^2/\omega_p^2)$ is the critical power for
relativistic self-focusing \cite{sun}, $P_c\simeq17(n_c/n_e)$\,GW, and
$r_e=e^2/m_ec^2$ is the classical electron radius.
\begin{figure} [!ht]
\centering{\includegraphics[width=9.8 cm]{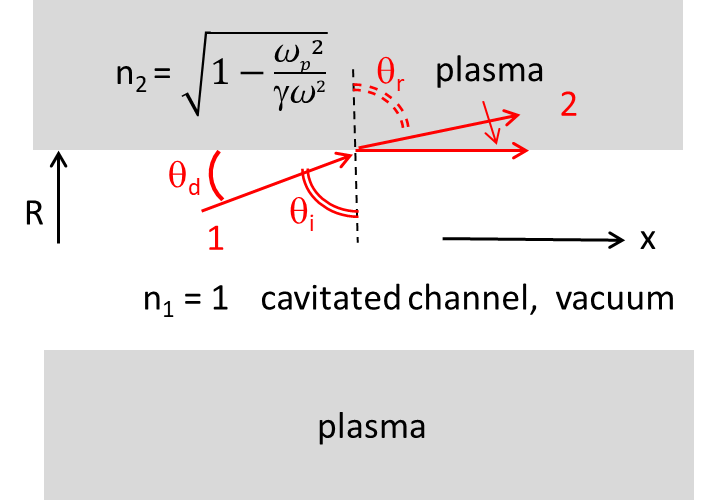}}
	\caption{Explanatory scheme for the self-trapped propagation of an intense
light pulse.}
	\label{fig3}
\end{figure}

We now present a simple, elegant derivation that explains matching
condition~(\ref{eq1}) for the cavern spot size. We start with standard
arguments usually used for a general tutorial explanation of self-focusing
in a nonlinear medium. For strong cavitation, the relativistically strong
electromagnetic fields in the laser focus produce an extremely high
charge-separation electrostatic field expelling electrons from the laser
axis and relativistically modify the electron mass such that the plasma acts
as a positive lens. We apply a geometric-optical consideration to the laser
pulse propagation in a piecewise-inhomogeneous medium with a cylindrical
channel (cavity) around the axis and plasma outside. In the cavity, light
naturally diffracts with a typical angle $\theta_d\simeq\lambda/\pi R$,
which defines an angle of incidence $\theta_i=\pi/2 -\theta_d$, as shown in
Fig.~\ref{fig3}. Snell's law $n_1\sin\theta_i=n_2\sin\theta_r$ governs the
behavior of light rays as they propagate across an interface between two
media (medium~1: vacuum with $n_1=1$; medium~2: plasma with $n_2=
\sqrt{1-\omega_p^2/\gamma\omega^2}$) and defines the condition $\theta_r=
\pi/2$ of the total internal reflection, which requires
\begin{equation}\label{eq3}
\theta_d^2\simeq\left(\frac{2c}{\omega R}\right)^2\simeq
\frac{\omega_p^2}{\gamma\omega^2}\simeq
\frac{\sqrt{2}\omega_p^2}{a_0\omega^2}
\end{equation}
for pulse propagation with an almost self-sustaining cavity radius. Here, we
use $\gamma=\sqrt{1+a_0^2/2}\simeq a_0/\sqrt{2}$. From Eq.~(\ref{eq3}), we
easily obtain Eq.~(\ref{eq1}), where $\alpha=2^{0.75}$, which is between the
values assumed in Refs.~ \cite{gordienko,Popov} ($\alpha=1.12$) and in
Refs.~\cite{Lobok,lu} ($\alpha=2$).
\begin{figure} [!ht]
\centering{\includegraphics[width=7.7 cm]{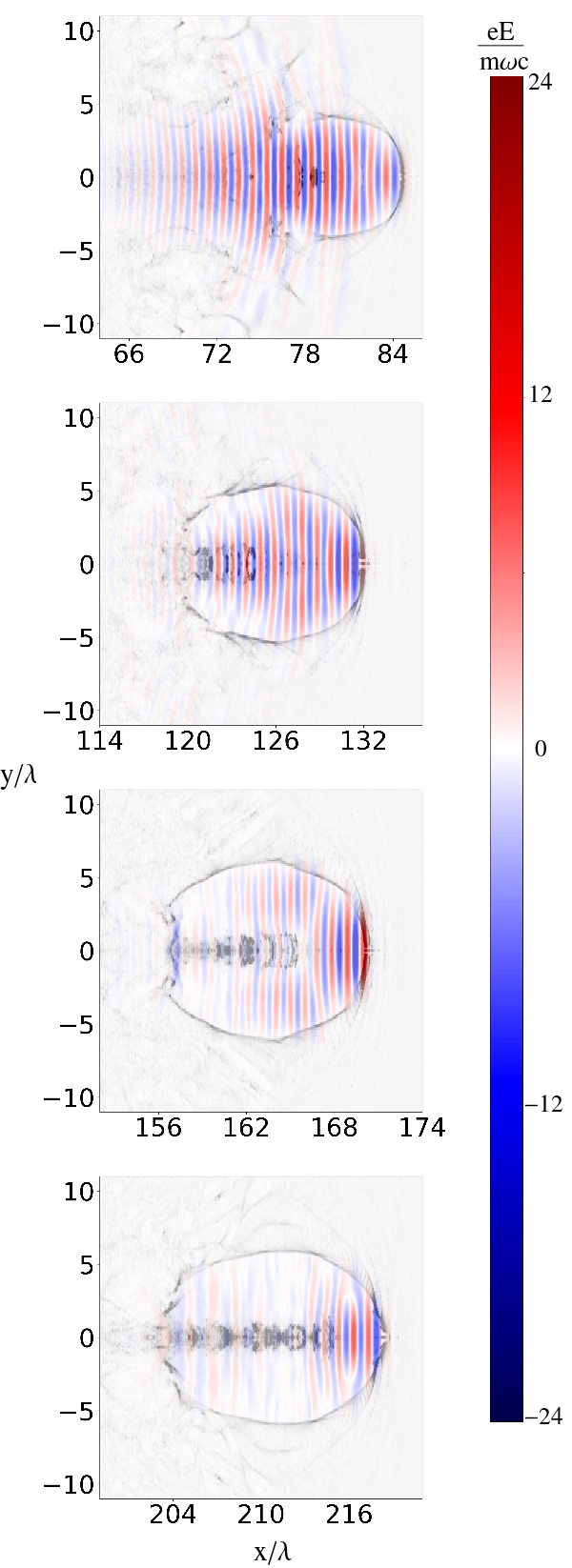}}
	\vspace*{-4mm}\caption{Laser pulse propagation for optimum
condition Eq.~(\ref{eq1}) ($a_0=24$, $R_L=2\lambda$, $n_e=0.1n_c$).}
	\label{fig4}
\end{figure}

Because of strong diffraction, radii smaller than given by Eqs.~(\ref{eq1})
and (\ref{eq2}) do not support self-channeled propagation of a laser beam,
while filamentation makes a wider laser pulse unstable. The necessary
condition for the absence of filamentation is that the laser pulse must
initially be sufficiently narrow. Filaments cannot appear only if the
laser pulse is not wider than the plasma skin length $R>(c/\omega_{pe})
\sqrt{a_0}$ (see Eq.~(\ref{eq1})). We note that condition~(\ref{eq1})
also corresponds to the force balance, where the ponderomotive force is
counterbalanced by the radial electrostatic focusing force due to the ion
cavity charge. By the above arguments, we can identify the propagation
regime with the matched pulse spot size given by Eq.~(\ref{eq1}) as the most
reliable for stable pulse propagation, which is expected to provide the
highest current of accelerated electrons because pulse propagation is stable
over a long distance. A regime of practically unchanged transverse laser
beam size (i.e., a self-trapping regime) is known in relativistic
self-focusing theory, which gives Eq.~(\ref{eq1}), where $\alpha=\sqrt{2}$
in the paraxial ray approximation with a simplified relativistic
nonlinearity \cite{walia}. We note that the condition of complete electron
cavitation immediately at the entrance of the light beam into the target
also requires the same pulse radius in accordance with the most advanced
theory, where self-focusing is associated with plasma nonlinearities due to
both relativistic electron mass variation and relativistic
charge displacement \cite{kovalev}.

Our simulations demonstrate stable low-dissipation pulse propagation through
many Rayleigh lengths, $X_R=\omega R_L^2/c$, in the form of self-trapped
light ($L>D$, $L\gg\lambda_p/2$), as shown in Fig.~\ref{fig4}. The laser
field is shown in red-blue colors. There is also the longitudinal
electrostatic field within laser cavities. It has a form of a single-cycle
field with the wavelength equal to the cavity length $L$. The single-cycle
electrostatic plasma field is strongly modulated because of the significant
electron charge of the electron microbunches in the laser field with a
$\lambda$-scale spacing. The longitudinal electrostatic field $E$ in the
laser cavity exceeds the Tajima--Dawson value ($E_{TD}=m_e\omega_pc/e$):
$E=4.5E_{TD}\simeq4.5$\,TV/m.

For the parameters used, the pulse propagation regime corresponds well to
condition~(\ref{eq1}). The formed laser cavity propagates ten Rayleigh
lengths ($\sim250\lambda$) for $a_0=24$, $R_L=2\lambda$, and $n_e=0.1n_c$
from entering the plasma to the distance where the pulse depletes. We found
that the estimate of the characteristic depletion scale $L_d\approx Ln_e/n_c
\simeq90\lambda$ from the effect of pulse depletion \cite{decker} agrees
with our simulation result similar to that in Ref.~\cite{lu}. This depletion
is clearly seen in Fig.~\ref{fig4} and is expressed as a smooth shrinking
of the pulse length to almost a single wavelength. Until the instant this
happens, a plasma cavity is well pronounced, as seen Fig.~\ref{fig4}, where
the relative electron density is represented as different gray levels
(darker denotes denser).
\begin{figure} [!ht]
\centering{\includegraphics[width=16 cm]{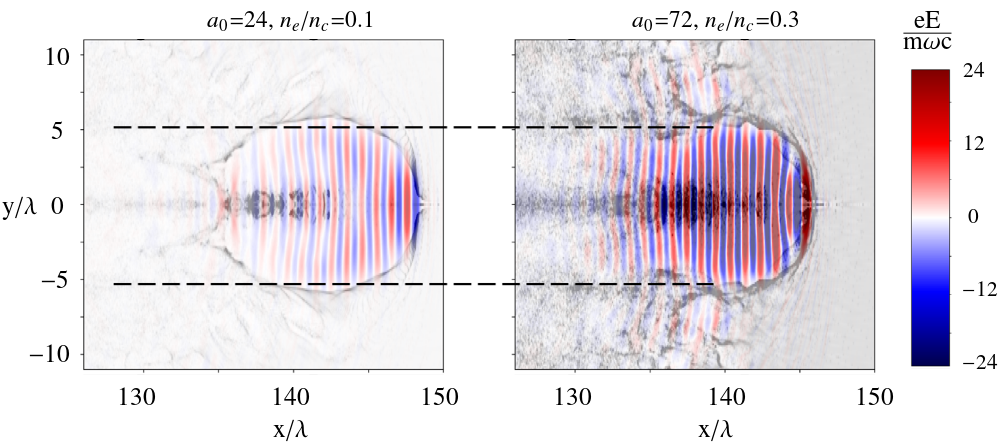}}
	\caption{Comparison of two self-trapping pulses with initial radius
$R_L=2\,\mu$m and amplitudes $a_0=24$ (left) and $a_0=72$ (right)
propagating in plasmas with the corresponding electron densities $0.1n_c$
and $0.3n_c$.}
	\label{fig5}
\end{figure}

\begin{figure} [!ht]
\centering{\includegraphics[width=16 cm]{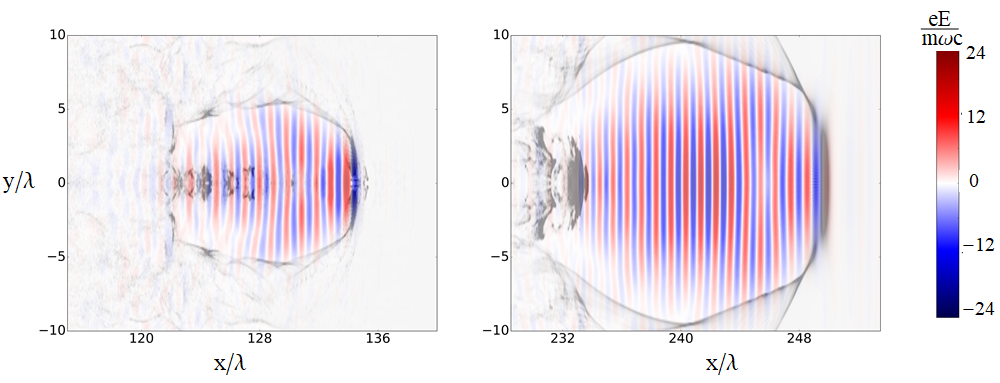}}
	\caption{Comparison of two self-trapping pulses with the amplitude
$a_0=24$ and initial radii $R_L=2\,\mu$m (left) and $R_L=4\,\mu$m (right)
propagating in plasmas with the corresponding electron densities $0.1n_c$
and $0.02 n_c$.}
	\label{fig6}
\end{figure}
Matching condition~(\ref{eq1}), which corresponds to the most stable laser
pulse propagation in the self-trapping regime and the maximum charge of
high-energy electrons was also checked in PIC simulations by varying the
laser field amplitude and laser hot spot size. Figure~\ref{fig5} clearly
demonstrates preservation of the laser pulse size with a constant ratio
$a_0/n_e$. If the laser focal spot size doubles from $R_L=2\,\mu$m to
$R_L=4\,\mu$m , then the size of the self-steady accelerating structure also
doubles and becomes $R=10\,\mu$m. The optimal plasma density correspondingly
decreases $\sim4$ to 5 times in accordance with Eq.~(\ref{eq1}), which
demonstrates the validity of this matching condition for another radius $R$.
This is illustrated in Fig.~\ref{fig6}.

\section{Electron acceleration}

In our simulations, we collect all the electrons escaping in the forward
direction behind the target in vacuum with energy greater than 30\,MeV. For
example, such electrons can be used to generate MeV gammas from a converter
target of high atomic charge number (see Fig.~\ref{fig1}). For each given
laser power, we performed several runs in the vicinity of Eq.~(\ref{eq1}) to
maximize the total charge $Q$ of high-energy electrons. Such maximization
also includes a proper choice of the target length $l\simeq2L_d$ to $3L_d$,
i.e., about the length of entire pulse depletion.
\begin{figure} [!ht]
\centering{\includegraphics[width=8.2 cm]{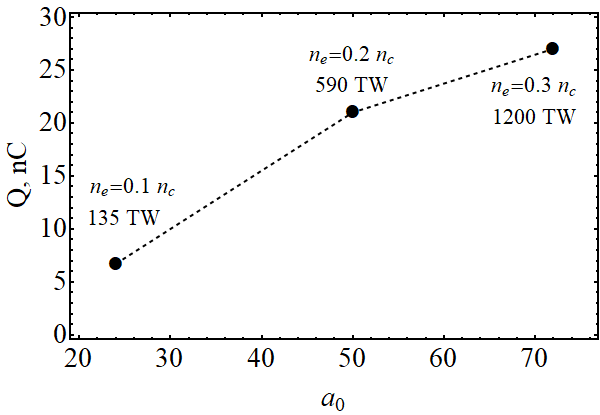}}
	\caption{The maximum total charge of electron bunches for electrons
accelerated in the laser self-trapping regime to energies $>30$\,MeV.}
	\label{fig7}
\end{figure}

The detected charges of high-energy electron bunches with energies greater
than 30\,MeV behind a target are shown in Fig.~\ref{fig7} for three values
of power. The electron acceleration scenario is quite different from the
bubble wakefield regime, which requires a laser pulse length less than the
plasma wavelength, $\lambda_p=2\pi c/\omega_{pe}$, where $\omega_{pe}$ is
the electron plasma frequency. The laser pulse does not excite a plasma wave
behind it. At the same time, the laser pulse is too short to excite plasma
waves driven by forward stimulated Raman scattering (FSRS) \cite{sprangle}.

\begin{figure} [!ht]
\centering{\includegraphics[width=12.8cm]{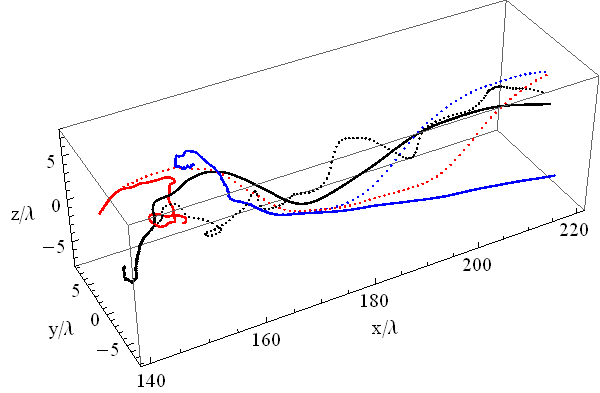}}
\caption{Comparison of how $10^{-4}$ perturbations affect the behavior of
several electron trajectories: the reference trajectories are shown by solid
lines and the perturbed trajectories are shown by dotted lines.}
\label{fig8}
\end{figure}

Ambient electrons flow into the accelerating cavity from its backside
similarly to the standard bubble wakefield case because laser field is
weakened at the pulse tail. These electrons enter strong laser and plasma
fields and accelerate quickly to the speed of light. They co-propagate
with the laser pulse and electrostatic field for a rather long time, but the
laser field is finally depleted during propagation, and the particle motion
becomes increasingly dominated by the longitudinal electrostatic field. The
pre-acceleration mechanism of the electrons could be called a specific kind
of direct laser acceleration enabling an effective loading of a large number
of particles into the accelerating electrostatic plasma field. An amplitude
modulation of such a single-cycle electrostatic field also contributes
stochastically to the occurrence of electron acceleration similarly to that
observed for FSRS \cite{bochkarev}.

The stochastic nature of electron loading can be proved by the following
simulation trick, which replaces the more complicated Lyapunov exponent
analysis \cite{bochkarev}. We compared electron trajectories with those
from an additional run in which we studied the system stability by adding
small momentum perturbations ($|\delta{\bf p}|/{\rm p}=10^{-4}$) for some
electrons at a given instant in the initial stage of electron
pre-acceleration, similarly to what was done in Ref.~\cite{sentoku}.
Figure~\ref{fig8} compares the electron trajectories with and without
perturbations and shows the significant difference in the final particle
trajectories, proving the stochastic nature of the electron acceleration.
\begin{figure} [!ht]
	\centering{\includegraphics[width=9.8cm]{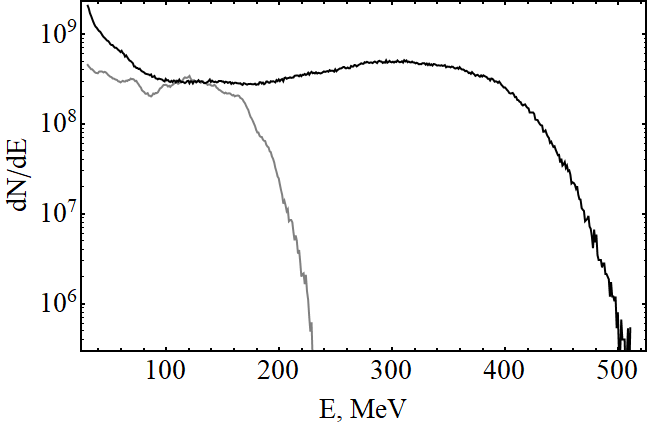}}
	\caption{Energy spectra of high-energy ($>30$\,MeV) electrons that left the
target: they correspond to the two laser powers in Fig.~\ref{fig5},
$P=135$\,TW (gray curve) and 1200\,TW (black curve).}
	\label{fig9}
\end{figure}

The spectra of accelerated electrons in the self-trapping regime from
low-density targets show their enrichment by high-energy particles. The
electron spectra have a well-pronounced energy plateau, as can be seen
in Fig.~\ref{fig9}. Such flattened distributions are formed from
accelerating electrons of modest energies and high-energy electrons that are
already dephased with respect to the electrostatic field and are
decelerating. Introducing an effective temperature $T_h$ as the ratio of the
total bunch energy to the total number of electrons in the bunch (see
Fig.~\ref{fig9}), we correspondingly derive $T_h=97$\,MeV and $T_h=210$\,MeV
for $P=130$\,TW and $P=1200$\,TW. The laser-to-electron beam energy
conversion efficiency is $\sim15$\% for all laser energies considered above.

\section{Nuclear emission from bremsstrahlung converter target}

Since the first proposals to use intense ultrashort laser pulses to trigger
nuclear reactions appeared \cite{bychenkov,ledingham}, this issue has been
overgrown with numerous original applications. Electron acceleration by a
laser-excited plasma wakefield have been used to produce gamma rays by
passing through high-Z material converters. This technique has distinct
advantages over direct laser irradiation of solid targets because the source
size is small. One of the first experiments on generating bremsstrahlung
gamma rays from laser wakefield acceleration was reported in
Ref.~\cite{edwards}.

For high photon energies beyond several MeV, generation by bremsstrahlung
is most attractive. It is typically produced when electrons accelerated in
low density are converted to high-energy photons in a high-Z material. The
gamma source size is rather small, which makes it potentially useful for
gamma-ray radiography with a high spatial resolution. Gamma-ray radiography
with a single shot is often very desirable. This allows the users to have
gamma-imaging with a high temporal resolution. Certainly, only
high-brightness sources can work in this way, and the highest possible
charge of the accelerated electrons is required.

Significant developments in the laser acceleration of electrons to energies
exceeding 100\,MeV have enabled tabletop photonuclear physics to be explored.
Such energetic electron beams have an array of applications in radioisotope
production, photofission induction, neutron generation, electron--positron
pair generation, and even possibly light meson production. The latter was
already demonstrated experimentally \cite{schumaker}. A laser-driven neutron
source has a very short temporal scale, which is favorable for applications
that use pulsed neutron sources, for example, fast neutron resonance
radiography \cite{mor}. In our numerical simulations (see Fig.~\ref{fig1}),
we use a high-charge electron bunch to irradiate high-Z material with the
aim to produce both gamma rays and photonuclear reactions with a high yield.

\subsection{Generation of gamma rays} \label{gamma}

An electron bunch accelerated by a short pulse in a laser target was used
for bremsstrahlung production of gamma rays from the second target (high-Z
converter target) placed immediately behind the laser target. Using the
Monte Carlo simulation tool GEANT4, we calculated the total yield of gamma
rays emitted from a Pt target (of natural composition,
$^{194}$Pt+$^{195}$Pt+$^{196}$Pt+$^{198}$Pt) of different thicknesses, the
gamma energy spectrum, and the angular distribution of gamma rays for a
100\,TW laser pulse.
\begin{figure} [!ht]
	\centering{\includegraphics[width=12.5cm]{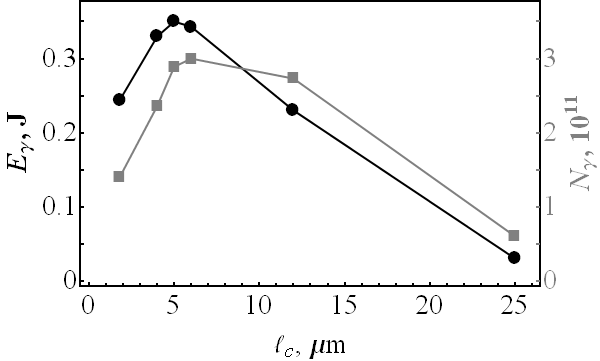}}
	\caption{Total yield (left) and energy (right) of gamma rays versus the
thickness $l_c$ of the Pt converter target for the laser--plasma
parameters $P=130$\,TW, $R_L=2\lambda$, and $n_e=0.1n_c$ corresponding to an
electron bunch with $Q\simeq7$\,nC and an average energy 100\,MeV.}
	\label{fig10}
\end{figure}

The total yield and energy of gamma photons leaving the converter target as
a function of its thickness are shown in Fig.~\ref{fig10}. The maximum gamma
yield (MGY) converter has a thickness $\sim5$ to 6\,mm, which somewhat
exceeds the electron stopping length of 100\,MeV particles in Pt
($\sim3$\,mm), i.e., such a converter target stops the electrons almost
entirely and is transparent for MeV gamma radiation. The maximum total yield
of MeV gamma radiation is as high as $3\times10^{11}$ photons, $\sim0.35$\,J.
Correspondingly, this gives $\sim8\%$ laser-to-gamma conversion efficiency,
which is much higher than 3D PIC--Monte Carlo simulations predict for
gamma bremsstrahlung production from ultraintense femtosecond
laser--solid interactions with front surface structures \cite{jianga} or
with a lengthy plasma corona \cite{lobok2017}. A gamma ray source based on
bremsstrahlung radiation generated by wakefield accelerated electrons in a
rarefied plasma \cite{cipiccia} is also noncompetitive relative to the
considered regime because the electron charge is significantly lower.

\begin{figure} [!ht]
	\centering{\includegraphics[width=7.5cm]{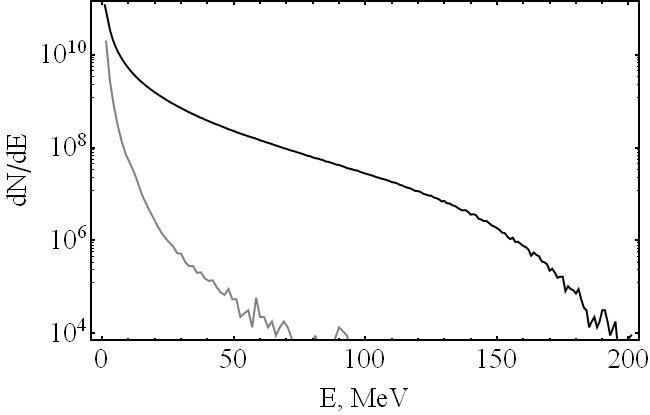} \hspace{1cm} \includegraphics[width=7.cm]{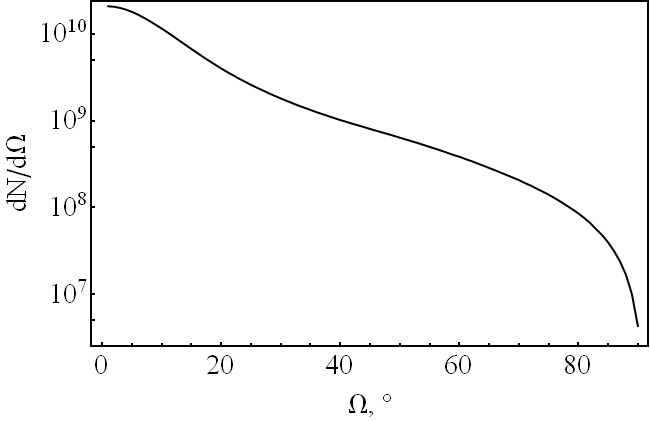}}
	\caption{Energy spectra (left panel) and angular distribution of gamma
rays generated from a 6\,mm thick Pt target in the forward (black) and
backward (gray) directions for the laser--plasma parameters $P=130$\,TW,
$R_L=2\lambda$, and $n_e=0.1n_c$ corresponding to an electron bunch with
$Q\simeq7$\,nC and average energy 100\,MeV. }
	\label{fig11}
\end{figure}
Figure \ref{fig11} illustrates a spectral distribution of the generated
gamma rays of multi-MeV energy with a strong dominance of their production
in the forward direction and the energy-integrated angular distribution of
gamma rays (left and right panels). The energy spectrum shows a
two-temperature distribution of gamma rays with
$T_{\gamma}\simeq2.5$\,MeV for MeV gamma rays and $T_{\gamma}\simeq25$\,MeV
for photons with energies exceeding 30\,MeV. The energy of high-temperature
gamma photons is $\sim0.1$\,J.

To estimate the gamma ray brightness, we use the duration and spot size of
gamma rays from the GEANT4 simulation. Both the duration and spot radius
depend weakly on the gamma energy but depend strongly on the converter
target thickness. For the MGY converter target of 5 to 6\,mm thickness, they
are 0.3\,ps and $60\,\mu$m. These values are much higher than the electron
beam duration ($\sim30$\,fs) and size ($\sim5\,\mu$m) often used to estimate
the gamma brightness (see, e.g.,~\cite{jianga}). For the corresponding
divergence $10^\circ$ in the simulation, we obtain
$\simeq5\times10^{17}$s$^{-1}$mrad$^{-2}$mm$^{-2}$(0.1\%BW)$^{-1}$ for the
10\,MeV gamma brightness. On the other hand, for the thinner target
(1.8\,mm thickness), which produces a gamma pulse of 100\,fs duration and
$17\,\mu$m spot radius, the 10\,MeV gamma brightness increases more than
an order of magnitude to $\sim10^{19}$s$^{-1}$mrad$^{-2}$mm$^{-2}$(0.1\%BW)$^{-1}$
despite the considerable reduction in the total number of generated gamma
photons (see Fig.~\ref{fig10}). The brightness of the 100\,MeV gamma source
is approximately an order of magnitude less than the above estimates.

\subsection{Photoproduction of electron--positron pairs}

\begin{figure} [!ht]
	\centering{\includegraphics[width=7.5cm]{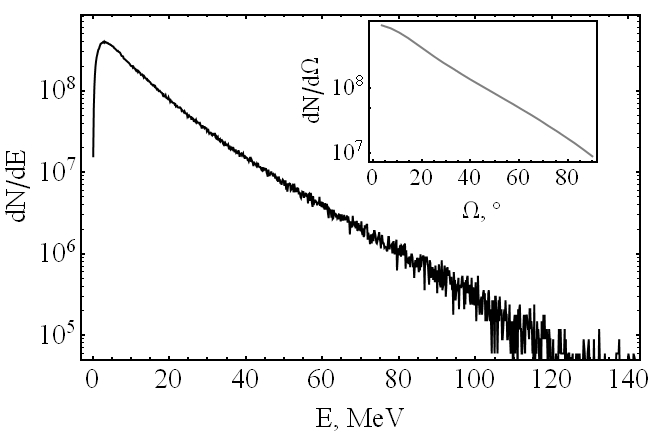} \hspace{1cm} \includegraphics[width=6.cm]{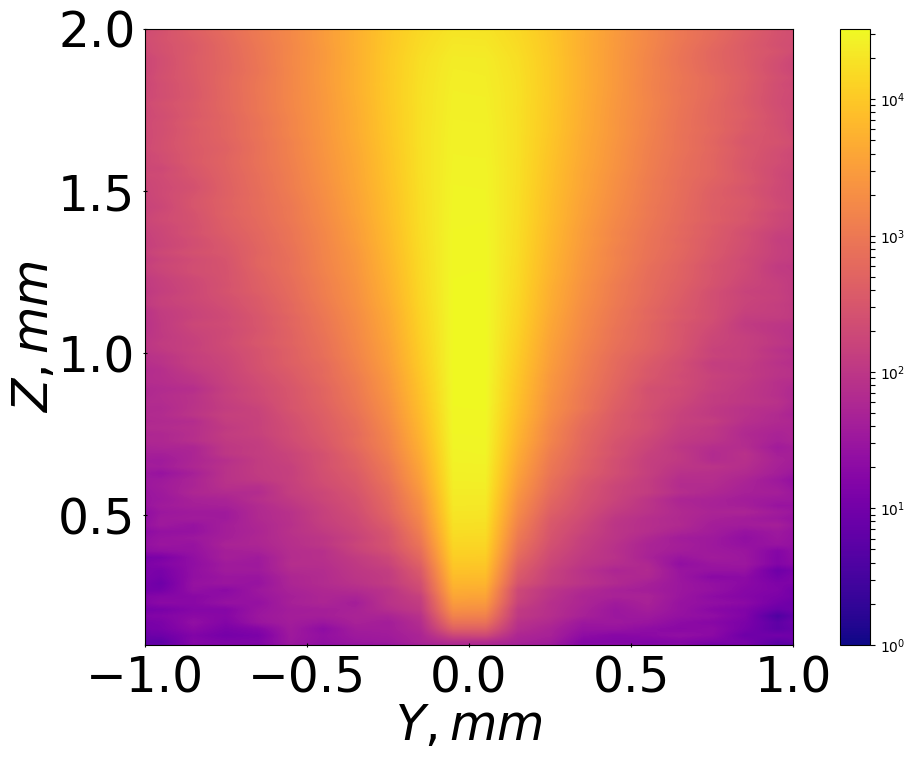}}
	\caption{Energy spectrum of positrons generated from a 6\,mm Pt target
(left panel) and positron flux in a 1.8\,mm Pt target (right panel) for the
laser--plasma parameters $P=130$\,TW, $R_L=2\lambda$, and $n_e=0.1n_c$
corresponding to an electron bunch with $Q\simeq7$\,nC and average energy
100\,MeV. The inset in left panel shows the angular distribution of
positrons generated from the 6\,mm Pt target.}
	\label{fig12}
\end{figure}

Our next GEANT4 simulations were performed to study electron--positron pair
production with a 100\,TW laser pulse. The electron beam was acting on the
same Pt slab of 6\,mm thickness, which also maximizes positron yield. The
positrons were considered when escaping from the backside of this target.
The thicker target gives fewer electrons and positrons escaping from it. The
bremsstrahlung emission of photons by electrons and the creation of
electron--positron pairs by photons are the dominant processes leading to
the generation of the detected electron--positron jet (see the right panel
in Fig.~\ref{fig12}).

We compared positron distributions for converter targets of different
thicknesses. A thickness increase results in angular blurring of generated
positrons. A better collimated positron jet results from a thin target
(shown inside a converter the right panel in Fig.~\ref{fig12}). The positron
jet angular spread increases from $\sim20^\circ$ for a 1.8\,mm Pt converter
target to $\sim35^\circ$ for a 6\,mm target and $\sim40^\circ$ for a 12\,mm
target. The spectrum of positrons from a 6\,mm Pt target is shown in
Fig.~\ref{fig12}. Simulations show that majority of positrons are produced
with an energy of the order of a few MeV, which corresponds well to a
J\"uttner--Synge distribution \cite{synge}. The total number of generated
positrons is $9\times10^9$, which is higher than observed in Ref.~\cite{sarri}.
But the energy was lower there because more energetic electrons were
generated in Ref.~\cite{sarri}. The 6\,mm thick Pt target produces the
maximum number of positrons. Because such a relatively thin target is unable
to stop all electrons from the injected electron beam, the total number of
electrons behind the converter target is an order of magnitude higher than
the number of positrons. Increasing the Pt target thickness to 24\,mm
results in an almost neutral electron--positron hot plasma jet behind the
converter with the total number of particles of two orders of magnitude less
than in the optimum thickness case. We note that in this case, there is a
large number of low-energy electrons appearing due to secondary processes
like Compton, Moller, and Bhabha scattering accounted for in the GEANT4 code.

\subsection{Photonuclear production of neutrons}

Photonuclear production of neutrons is a widely discussed topic in nuclear
applications of short-pulse intense lasers. One scheme is to irradiate the
converter with high-energy electron beams from laser wakefield acceleration
\cite{reed2007,jiao}. We discuss this here using the self-trapping regime.
Such a regime enables generation of gamma rays with energies exceeding
10\,MeV. This gamma energy range is well suited for photonuclear neutron
production through the giant dipole resonance (GDR). The isovector GDR is
known as the fundamental collective nuclear excitation \cite{harakeh}. The
GDR can be understood macroscopically as a bulk nuclear vibration where
protons with isospin T$_3$=1/2 and neutrons with isospin T$_3$=-1/2
oscillate coherently in opposite directions. Coherent excitation makes a
sufficiently large cross section. In general, the photonuclear cross section
is smaller than typical nuclear cross sections because of the
electromagnetic nature of the interaction. Nevertheless, at resonance
energy, it is comparable in order of magnitude to the geometric nuclear
cross section, which well compensates the weak electromagnetic interaction.
It is larger than the most (n, $\gamma$) cross sections.

\begin{figure} [!ht]
	\centering{\includegraphics[width=8.cm]{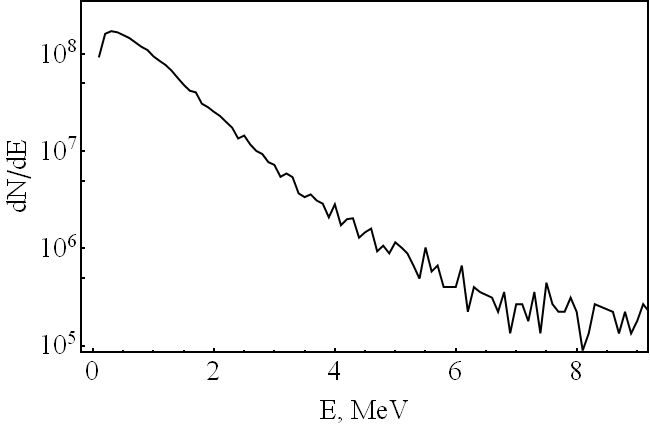} \hspace{1cm} \includegraphics[width=6.cm]{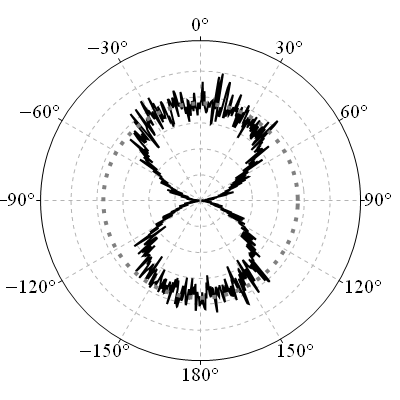}}
	\caption{Energy spectrum (left panel) and angular distribution (right
panel) of neutrons generated from a 12\,mm thick Pt target outside the
target in the forward and backward directions (in a $108^\circ$ angle in
both cases) for the same laser--plasma parameters as in Fig.~\ref{fig11}.}
	\label{fig13}
\end{figure}

Photonuclear cross sections typically have broad peaks with relatively high
values in the GDR energy region 10 to 20\,MeV. This is the case with the
gamma source presented in Sec.~\ref{gamma} (see Fig.~\ref{fig8}). The
photonuclear reactions at the GDR region are mainly ($\gamma$, 2n) and to
a lesser extent ($\gamma$, n), although Geant4 (version 10.5), which is
well suited for selected GDR cross sections, also simulates a variety of
reactions: ($\gamma$, n), ($\gamma$, 2n), ($\gamma$, np), ($\gamma$, 3n),
and so on. We note that the secondary gamma radiation emitted by excited
nuclei are also taken into account in Geant4 simulations.

The results for photonuclear neutron production with a 130\,TW laser are
shown in Fig.~\ref{fig13}. The maximum number of generated neutrons from a
Pt target with a 12\,mm thickness reaches $2{\times}10^8$ particles. It
roughly corresponds to a $5{\times}10^{-3}$ electron--neutron conversion
efficiency \cite{barber1959neutron}. Using the 6\,mm Pt target (which
maximizes gamma and positron yields) cuts the neutron number in half
compared with the considered 12\,mm Pt target. The thicker converter target
is needed for neutron production because the photonuclear (n, $\gamma$)
cross section is much less than the positron production cross section and a
longer propagation distance is hence required. The generated neutrons have
an almost isotropic distribution and exponential-like spectrum with a
temperature $\sim1$\,MeV. Because we collect only particles moving in
forward and backward directions outside the converter target (in a
$108^\circ$ angle), the angular distribution in Fig.~\ref{fig13} (right
panel) does not contain cross-side generated neutrons. It shows some small
forward--backward anisotropy. The number of forward-produced neutrons
exceeds the number of backward ones by approximately 10\%.

\subsection{Photoproduction of pions}

Pions are light mesons (consisting of quarks and antiquarks) with a rest
mass of 140\,MeV. They are a key decay product in high-energy particle
physics experiments. Despite having a larger rest mass than muons, they
require less energy (threshold energy is 140\,MeV) to appear with
gamma-produced (virtual) quark--antiquark pairs in nuclei than for
electromagnetic production of muon--antimuon pairs (the required energy is
212\,MeV).

\begin{figure} [!ht]
	\centering{\includegraphics[width=9.8cm]{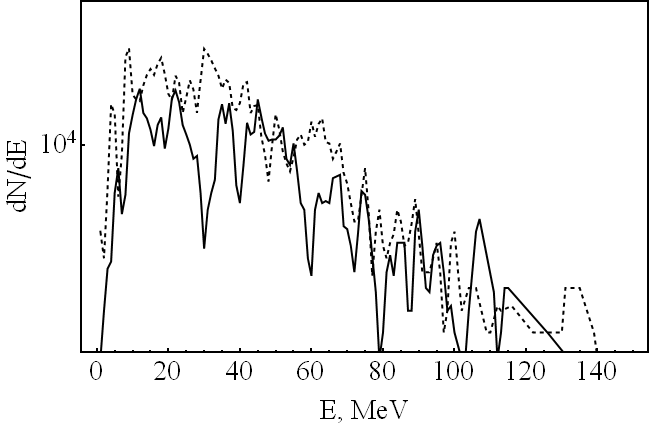}}
\caption{Spectra of photonuclear generated $\pi^+$ (solid line) and
$\pi^-$ (dashed line) pions for a 1200\,TW laser pulse.}
\label{fig14}
\end{figure}

For gamma energies up to 1\,GeV, there are two main channels of pion
generation: excitation of baryon resonances and direct (nonresonant) single
pion production, i.e., due to quark extraction from the nucleus (neutron or
proton). The direct two-body channel $\gamma N\to\pi N$ dominates near a
threshold up to a gamma energy of 200 to 300\,MeV. For higher energies, the
photon wavelength becomes comparable to the nucleon radius, and photons
mainly interact with single nucleons by exciting baryon resonances. The
excitation of the three baryon resonances $\Delta(1232)$, $N*(1520)$,
and $N*(1680)$ is the most important. The lifetime of these resonances are
less than $10^{-23}$\,s (e.g., the lifetime of the $\Delta(1232)$ resonance
is $5.6{\times}10^{-24}$\,s). The resonances commonly decay with pion
production. In the case of a 135\,TW laser pulse, the maximum electron beam
energy is only slightly above the pion production threshold (see
Fig.~\ref{fig7}), and this results in a small number of photons with
near-threshold energies of 140\,MeV. Therefore, our simulation predicts the
generation of only about 150 pions of different species ($\pi^+$, $\pi^-$,
and $\pi^0$). This number is comparable to the experimental results in
Ref.~\cite{schumaker}, where only direct pion production was detected. The
total number of generated pions dramatically increases for a 1200\,TW laser
pulse. It reaches $\sim10^6$ for each pion species). The spectra of $\pi^+$
and $\pi^-$ pions are presented in Fig.~\ref{fig14}.

The lifetime of the $\pi^{0}$ pion is $8.4{\times}10^{-17}$\,s and it decays
(with a probability 0.98823) to two gamma photons. The lifetime of the
$\pi^{\pm}$ pions is 26\,ns, and they decay (with a probability 0.999877) to
the corresponding $\mu^{\pm}$ muons. These decays are the main channel of
muon production in our case. For direct muon production due to
photon--nucleus interaction, the higher photon energies and correspondingly
higher electron beam energies are required. This agrees with the prediction
in Ref.~\cite{rao}, where direct muon production was numerically observed
with a multi-GeV electron beam from laser wakefield acceleration.

\section{Conclusions}

We have performed 3D PIC simulations using the code VSim to model electron
acceleration in low-density targets with lasers of intensities from 0.8 to
$7\times 10^{21}$ and powers from 135 to 1200\,TW subsequently complemented
by Monte Carlo simulations using the code GEANT4 to model gamma, positron,
and photonuclear particle production. We considered the self-trapping regime
of relativistic laser channeling in a near-critical plasma, which enables
stable laser pulse propagation over many Rayleigh lengths. This regime is
the most suitable for photonuclear reactions because it can provide the
maximum total charge (multi-nC) of electrons accelerated to a hundred MeV
with current short-pulse intense lasers. For PW class lasers, the electron
bunch charge can be as high as several tens of nC.

The considered self-trapping regime of laser pulse propagation corresponds
to the matched pulse spot size given by Eq.~(\ref{eq1}). This is a regime of
practically unchanged transverse laser beam size. The corresponding
nonlinear structure looks like an empty cavity of pulse length with an
electrostatic field filled with a laser field (a ``laser bullet"). The
considered relativistic laser pulse satisfies the condition of complete
electron cavitation immediately at the entrance of the light into the
target, which supports almost the same pulse radius over entire propagation
length until pulse depletion in accordance with the most advanced theory,
where self-focusing is associated with plasma nonlinearities due to both
relativistic electron mass variation and relativistic charge displacement
\cite{kovalev}. Acceleration of electrons in laser bullet occurs as a
combination of direct laser acceleration and electrostatic wake acceleration
with a stochastic feature. The electron energy spectra show a
well-pronounced plateau of the width $\Delta E\sim E\sim100$ to 300\,MeV
(see Fig.~\ref{fig9}).

One possible application of the proposed acceleration scheme is to generate
gamma quanta with characteristic energies up to tens of MeV via
bremsstrahlung emission by electrons passing through a heavy metal converter
plate placed behind the laser target. Such gamma radiation can be used for
radiography of dense samples a few tens of centimeters thick, which is in
great demand for materials science and security inspections. With a 130\,TW
30\,fs laser pulse, for example, we obtained an electron--photon conversion
efficiency at the level of 5\% for $\sim2.5$\,MeV gamma photons and 2\% for
$\sim25$\,MeV gamma photons. The estimated brightness of 10\,MeV gamma rays
is $\sim10^{19}$s$^{-1}$mrad$^{-2}$mm$^{-2}$(0.1\%BW)$^{-1}$. We estimated
the photo-production yield of neutrons, positrons, and pions, which is also
benefited by the high charge of the generated electron beam in the
self-trapping regime of the laser pulse. While the low repetition rate of
high-energy laser facilities has limited the development of photo-nuclear
sources, several multi-Joule and multi-Hz laser systems will become
available over the next five years. Our study highlights the potential of
high-repetition rate experiments by demonstrating the advantage of
near-critical gas targets.

\section{Acknowledgments}
	
This work was supported by the Russian Science Foundation (Grant
No.~17-12-01283).

\end{document}